\documentstyle[aps,epsfig]{revtex}
\newcommand{\be}{\begin{equation}}
\newcommand{\ee}{\end{equation}}
\newcommand{\kl}{{kagom\'e }}
\begin{document}
\title{Dirac, Anderson, and Goldstone on the Kagom\'e}
\author{M. B. Hastings}
\address{Physics Department, Jadwin Hall\\
Princeton, NJ 08544\\
hastings@feynman.princeton.edu}
\maketitle
\begin{abstract}
We show that there exists a long-range RVB state for the \kl lattice
spin-1/2 Heisenberg antiferromagnet
for which the spinons have a massless Dirac spectrum.  By considering various
perturbations of the RVB state which give mass to the fermions by
breaking a symmetry, we are able to describe a wide-ranging class of known 
states on the \kl lattice, including
spin-Peierls solid and chiral spin liquid states.  
Using an RG treatment of fluctuations about the RVB state, we propose
yet a different symmetry breaking pattern and show how collective excitations 
about this state account for the gapless singlet modes seen experimentally and
numerically.  We make further comparison with numerics for Chern
numbers, dimer-dimer correlation functions, the triplet gap, and other 
quantities.  To
accomplish these calculations, we propose a variant of the $SU(N)$ theory 
which enables us to include many of the effects of Gutzwiller projection at 
the mean-field level.
\end{abstract}
\section{Introduction}
The spin 1/2 Heisenberg antiferromagnet on the \kl lattice
is a good candidate for a
two-dimensional quantum system with a spin disordered ground state\cite{zengel}.
While it appears that on square\cite{square} and 
triangular\cite{triangle,triangle2,triangle3}
lattices an antiferrmomagnet will acquire N\'eel order, on the \kl lattice 
strong numerical evidence has accumulated that the system is spin disordered,
as seen by the existence of a gap to triplet excitations and through
consideration of the spectra of finite size samples\cite{numerics2}.
Numerically, one finds a continuum of low energy states below the
triplet gap\cite{numerics4}.  
The continuum of low energy excitations provides a great
puzzle to theory in the absence of an obvious broken symmetry.

There are good experimental realizations of \kl systems, 
despite the presence of additional couplings, including the jarosites
and $SrCrGaO$.  While in iron jarosites\cite{iron,iron2}, these 
additional couplings
produce long-range order, in deuteronium jarosite\cite{jaros} and 
$SrCrGaO$\cite{scgo} no long range order is seen.  Additionally, in $SrCrGaO$
a quadratic specific heat and very weak field dependence of the specific 
heat\cite{ramirez} are in agreement with the picture of
a continuum of low-energy singlet excitations seen in numerics, suggesting
that the latter two compounds provide good realizations of the \kl 
antiferromagnet.

Given the lack of spin order, RVB ideas seem natural for this system,
and indeed have stimulated much theoretical work on the system.
Large $N$ calculations based on $SU(N)$
have been used to suggest a spin-Peierls state\cite{mz}.
Calculations based on $Sp(2N)$ have suggested
a phase with deconfined, gapped, bosonic spinons\cite{subir}.
Chiral states have also been proposed\cite{mz}, but do not
account for the excitation spectrum and also are in disagreement
with the rapid decay of chirality-chirality correlation functions seen in
numerics.  States with BCS pairing have been suggested but again
do not account for the excitation spectrum; due to
the non-bipartite nature of the \kl lattice, these states are
{\it not} equivalent to flux states\cite{hsus}.
In addition to the long-range states,
short-range RVB states
based on a reduced Hilbert space of 
dimers\cite{srrvb,srrvb2,srrvb3} have also been considered and
provide
some explanation for the gapless continuum.

An RVB state on the \kl lattice would be particularly attractive, given
the intensive work on RVB states on the square lattice\cite{rvb}, especially
in connection with high-$T_c$ materials\cite{htc}.  In the
absence of doping, the square magnet eventually acquires N\'eel order and
the spinon excitations disappear from the system.
Since the \kl lattice does not acquire N\'eel order, it could be
a very important model for a spin liquid or spin solid state.

The idea behind the present approach is a to start with a
long-range RVB treatment of the \kl lattice and consider various
ways of gapping the spinon excitation spectrum.
We will first construct a ``parent state" which
will be the best RVB state that does not break time-reversal symmetry
or any lattice symmetry.  

We will then demonstrate an interesting massless
Dirac structure for this state.  
Various other known RVB states can be obtained by perturbing the
parent state, lowering the symmetry and giving mass to the Dirac particles, so 
that the parent state unifies a wide class of states.
Physically, we expect that the system will
attempt to give mass to the Dirac particles and open a gap, 
picking out one of these other lower symmetry states. 
We will discuss the symmetry breaking through a renormalization group
treatment.  We will obtain some kind of spin solid state, and
some low energy Goldstone and gauge excitations, which we will
argue provide the low energy degrees of freedom seen experimentally.

The RVB states can be thought of by a decoupling procedure, in
which we decompose
spin-1 operators into pairs of spin-1/2 operators.  
Take a Hamiltonian
\be
\label{sham}
H=\sum_{<i,j>} J \vec S_i\cdot \vec S_j
\ee
where the sum extends over neighboring sites $i,j$.

Introduce the spinon fields $\psi^{\dagger}_{a}(i),\psi_{a}(i)$,
where $a=u,d$ labels up and down spinon fields.  We can then introduce
a Hubbard-Stratonovich field $t_{ij}=t^{\dagger}_{ji}$ such that 
\be 
\label{mft1}
H=\sum_{<i,j>} \Bigl(\psi^{\dagger}_a(i) t_{ij} \psi_a(j)+h.c.\Bigr)+
\frac{2}{J}\sum_{<i,j>}|t_{ij}|^2
\ee
By taking a mean-field in $t$, minimizing the total energy
of the fermions and the Hubbard-Stratonovich field, we obtain an RVB state.  
One must at some point project results onto the physical space in which
each site is singly occupied.  

Later, we will find this projection to be extremely important.  In
the absence of projection, the ideal mean-field state is almost always found
by taking a dimer covering of the lattice\cite{dimerize}, 
with $t_{ij}$ nonvanishing only on
the given dimers.  Projection can stabilize RVB states, so although
our first calculations will ignore the effects of projection, in
a naive mean-field, we will later discuss a projected mean-field
that includes some of the essential effects of projection.

We will then have to proceed beyond mean-field solutions.  We will
consider a functional integral with fields $\psi(i)$ and $t_{ij}$,
fluctuating about a saddle-point of the action.  There are a large
number of possible fluctuations in $t_{ij}$, including a set of pure
gauge fluctuations, as well as a set of gauge fields.  Most
other fluctuations can be ignored because they do not contribute to
the low-energy dynamics. 
However, there will be a particular set of fluctuations in $t_{ij}$
that produce a mass for the fermion field.  Although these fluctuations are not
gapless, we will retain these 
fluctuations due to their impact on the low energy dynamics of the
fermion field.  We will see using a renormalization group that the
effective action of these fields can differ greatly from that suggested
by the mean-field.

To outline the paper, we will first describe the parent state, and then
discuss how to perturb the parent state to obtain other proposed RVB
states.  Then we will discuss naive and projected mean-field theory
treatments of these states.  We will the proceed to a field theoretic
treatment of fluctuations about the mean-field and a renormalization
group that will suggest one particular symmetry breaking pattern.
We will discuss the pseudo-Goldstone and gauge modes that arise from
this symmetry breaking, and the mechanism that ultimately gives them
a very small energy gap.

Next, we proceed to a discussion of finite system size effects as a first
step in comparison with numerics.  These effects lead to an additional
flux for odd system sizes which leads to a nonvanishing
Chern number for odd system sizes.
We will then compare the low energy bosonic modes from the 
field theory to the low energy singlet modes found in numerical calculations
as well as checking dimer-dimer correlation functions and many-body
density of states.
\section{The Parent RVB State}
Although the short range RVB calculations provide one starting point for
the \kl lattice antiferromagnet, we will be interested in looking at
long range states instead.  Certainly, the short range RVB calculations
themselves suggest that long range antiferromagnetic correlations are
important; the variational energy of these states improve when second
neighbor dimers are included.  Further, while the \kl lattice has
a gap to triplet excitations, this gap is about an order of magnitude
smaller than $J$; from a short range calculation one might expect a
gap of order $J$ as that is the energy to break a dimer.  However,
Mila has suggested that within a short-range state it is possible
to have a triplet gap much smaller than $J$\cite{srrvb3}, so
the small triplet gap does not necessarily rule against a short-range
state.

The best RVB state on the \kl lattice antiferromagnet is a
chiral spin liquid\cite{mz}.  A similar chiral state\cite{ky,kl} was
obtained using a hardcore boson representation of the spins and transmuting
the statistics from bosonic to fermionic using a Chern-Simons field.
However, numerical calculations\cite{numerics1} do not support a large
chirality-chirality correlation function or expectation value of the
chirality operator, which would seem to rule these states out.  So,
we will look for the best RVB state that is not chiral.

Assuming that we are looking for a long-range RVB state in which
all $t_{ij}$ have the same magnitude, the only choice we have is how
much flux to put into the system.  The state we choose involves putting
$\pi$ flux through the hexagons, and no flux through the triangles.
This state offers a better mean field energy than any other nonchiral
RVB state, including the state with no flux through the system at all.

The unit cell of the \kl lattice consists of three sites on a triangle.
Once we add flux to the system, the unit cell doubles, and requires six
sites on two triangles.  We will find it convenient to double the
unit cell again, to twelve sites, including six sites on a
hexagon and the six sites which neighbor the hexagon.  This
cell is shown in figure 1.  The \kl lattice is made up of a triangular
lattice of these 12-site unit cells.  We have numbered the points in
the cell for later reference.

For now, let us assume that we pick $J$ such that $|t|=1$ within our
RVB state.  Then, the band structure for our RVB state is shown in
figure 2 scanning along the given line of momenta in the Brillouin zone.  
There is
a degeneracy of states: the bottom
line in the figure actually consists of four bands, while the other
four lines in the figure consist of two bands each, providing a total
of twelve bands.  At $(0,0)$, four bands meet at energy less than zero and
another four meet at positive energy. Near this point the
spectrum becomes relativistic.  

The particles occupy the lowest six bands of the system, meaning that
where the bands meet the spectrum becomes gapless.  The system
can gain energy by perturbing about our given RVB state; we expect
that the greatest gain in energy comes from opening a gap.  For this
reason, we will study the Dirac point and look at possible
perturbations to the Dirac equation.

At the Dirac point, the Schr\"odinger equation for the fermions becomes
\be
\label{dirac}
E\psi=v_f (\alpha_x k_x+\alpha_y k_y) \psi
\ee
where $\psi$ is a four component spinor, and the matrices
$\alpha_x,\alpha_y$ are anti-commuting $\alpha$-matrices.
The particular basis chosen for $\psi$ and for 
the $\alpha$-matrices is unimportant.
We find by explicit computation that
\be
\label{vfeq}
v_f=(0.408248...)|t|
\ee

Given the Dirac equation (\ref{dirac}), we would like to consider the effect
of perturbations $\delta t_{ij}$ on the low energy structure.  This
analysis will enable us to focus on those fluctuations in $t_{ij}$
which have the greatest impact on the low-energy dynamics and which
must be kept when we proceed to a field-theory treatment of fluctuations.  

From equation (\ref{mft1}), the system pays an energy cost equal
to $\frac{1}{J}|\delta t_{ij}|^2$, but it can gain energy by
opening a gap for the Dirac particles.   As a result, we look for the
perturbations which open the greatest gap for the Dirac particles for
given $|\delta t_{ij}|^2$.
Let us first proceed algebraically, considering possible
perturbations to the Dirac equation which will open a gap, and only then
ask how to obtain these perturbations from $\delta t_{ij}$.

A given perturbation $\delta t_{ij}$
will perturb the Dirac equation to
\be
E\psi=\Bigl(v_f (\alpha_x k_x+\alpha_y k_y)+M\Bigr) \psi
\ee
where $M$ is some matrix, the projection of $\delta t_{ij}$ onto the space of 
the four states at the Dirac point.  It may be
shown that the perturbation $M$ will be most efficient
at opening a gap when it anti-commutes with $\alpha_x,\alpha_y$.
By efficient, we mean that we wish to maximize the gap for
given ${\rm Tr}(M^2)$, as a first step to maximizing the gap for
given $|\delta t_{ij}|^2$.
Since there is only a sixteen dimensional space of matrices $M$, we
can easily characterize all matrices that have the
needed anti-commutation property; it is a four dimensional
vector space.

We will write three of the perturbations as matrices $M_i$, for i=1,2,3.
In terms of $\alpha$-matrices, they will be $M_1=\alpha_z, M_2=\beta, 
M_3=\beta\alpha_x \alpha_y\alpha_z$.  These perturbations anti-commute
with each other; in fact, one can make a change in basis in the
Dirac equation which leaves
$\alpha_x,\alpha_y$ unchanged, but produces continuous $O(3)$
rotations in the space of $M_1,M_2,M_3$.  This continuous symmetry
is only valid at low energy; it will be broken to a discrete symmetry
by lattice effects as discussed below.
By taking $M=\sum_i m_i M_i$, for some numbers $m_1,m_2,m_3$,
we open a gap equal to $\sqrt{\sum_i (m_i^2)}$.  We will refer to
these as nonchiral mass terms.

The fourth perturbation is of a different sort.  It is $M=m_c M_c$ with
$M_c=i \alpha_x \alpha_y$  (here, c stands for chiral and
we will refer to this as a chiral mass term).
This perturbation breaks parity and time-reversal symmetry.  
$M_c$ commutes with $M_1,M_2,M_3$.  As mentioned above, we are
interested in the most efficient way for the system to open a gap; since
$M_c$ does not anti-commute with $M_i$, it is most efficient for the
system to take either purely chiral mass or purely nonchiral mass, so
that $m_i=0$ or $m_c=0$.

Next, we would like to ask what perturbations
in the $t_{ij}$ will produce the desired mass matrix $M$.  We will
find that to produce $M_i$ requires dimerizing the system by making
the magnitudes of the $|t_{ij}|$ non-uniform; to produce $M_c$ requires
adding additional fluxes to the system.
Clearly,
there is a large degeneracy here, as there is a only 16 dimensional space of
matrices $M$ while in a given unit cell there is a 48 dimensional space
of perturbations to $t_{ij}$.  While some of the
degeneracy is due to the large number of possible gauge transforms on
$t_{ij}$, this does not completely alleviate the problem.
 Again, the question of efficiency
becomes important: for each $M$, there is a class of $t_{ij}$ which produce
the desired perturbation, but only one element in the class minimizes
$\sum_{i,j} |t_{ij}|^2$. 
We will find one unique perturbation
in $t_{ij}$ (up to arbitrariness in gauge) which produces the desired
mass matrix.

The perturbation that we will pick for $M_1$ is shown in figure 3.
One can see that all the horizontal bonds have been either decreased
or increased in strength, such that along a horizontal line the 
bonds alternate in strength while horizontal bonds which are
in a vertical column all have the same strength.

This perturbation is a dimerization of the $t_{ij}$.  The spins
form singlets most strongly across the largest $t_{ij}$, so that
dimerization of $t_{ij}$ tends to produce a spin solid and changes
the long-range RVB to a short range set of singlets.

This mass term breaks rotational and translational symmetry of
the lattice.  We will pick $M_2$ and $M_3$ to be lattice rotations
of $M_1$.  The continuous symmetry of the Dirac equation will then
be broken at the lattice level to a discrete symmetry of permutations
of $m_1,m_2,m_3$ under lattice rotation, while
lattice translations change the sign of any 2 of the 3 
mass terms $m_1,m_2,m_3$.
We will find later that while we obtain symmetry breaking and
produce a mass, the discrete nature of the lattice group will leave
us with only pseudo-Goldstone modes.

There are two mass terms which are symmetric under rotations.
They are 
\be
M_{12}=\frac{M_1+M_2+M_3}{\sqrt{3}}
\ee
\be
M_{6}=-M_{12}=\frac{-M_1-M_2-M_3}{\sqrt{3}}
\ee
where 12 denotes the fact that the $t_{ij}$ are strongest on the
12-site loop surrounding the unit cell, while 6 denotes the fact that
the bonds are strongest on the hexagons and triangles.  We show the
perturbation to $t_{ij}$ to produce $M_{12}$ in figure 4.

To estimate dimerization later, it will be useful to know connect
the change in $t$ to the eigenvalues of the mass matrix that arises.
One finds that if the bonds on the 12-site loop are increased by
$\delta t$, while those on the hexagons and triangles are decreased
by the same amount, then one produces a term $m_{12}M_{12}$ in
the Dirac equation with $m_{12}=1.57735.... \delta t$.

For calculations later, it will be convenient to transform to a
basis of gamma matrices.  Defining $\gamma_t=\beta$, $\gamma_i=\beta\alpha_i$,
we find that $M_1=\gamma_3$, $M_2=1$, $M_3=i \gamma_5$, while
$M_c=i\gamma_t\gamma_x\gamma_y$.

It is interesting to compare to above characterization of possible
perturbations in terms of gamma matrices to the situation in the
$\pi$-flux phase on the square lattice, where there is again a
Dirac spectrum, and again various mass terms can be introduced\cite{mudry1}.  
We find that we are able to introduce one nonchiral mass term
by dimerizing the horizontal bonds of the square lattice, so that 
the horizontal bonds alternate in strength as one moves horizontally
along the lattice; another mass term can be introduced by dimerizing
the vertical bonds.  In the limit of extreme dimerization, these
states correspond to short-range RVB states in which the dimers
are stacked on top of each other, and
all lie either horizontally or vertically\cite{sqd}.  By taking sums of these
two mass terms, we can produce a short range state in which dimers resonate
around a square.  The final nonchiral mass term
can be obtained by placing an on-site potential on one sublattice of
the square lattice; this corresponds to introducing N\'eel order into
the system\cite{hsu}.
Due to the highly frustrated nature of the
\kl lattice, in this paper
we will not have any such terms involving introducing on-site
potentials.

For the square lattice, the chiral mass term can also be introduced.
It requires adding additional couplings to the system, which connect
diagonally across a given plaquette, and
then inserting $\pi/2$ flux through the triangle that is formed when a particle
traverses two sides of a plaquette and then crosses the plaquette on
the diagonal\cite{muthu}.

\section{Connection to Other Valence Bond States}
From the parent state, it is possible to continuously connect to
other possible valence bond states, using the mass terms
we have found above.  Let us first consider the chiral spin
liquid and then the spin-Peierls solid of Marston and Zeng\cite{mz}.

Let us consider a state such that the flux through
each triangle is equal to $\theta$, and the flux through each hexagon
is equal to $\pi-2\theta$.   Then, as $\theta$ varies from $0$ to
$\frac{\pi}{2}$ we continuously deform from the parent state to the chiral 
spin liquid.  By looking at how $t_{ij}$ changes along this deformation,
and then projecting this change onto the Dirac point, we find that for 
small $\theta$, the perturbation exactly produces the chiral mass term
$M_c$!
Let us note that at $\theta=\frac{\pi}{4}$ there is a highly interesting
band structure, discussed in the Appendix, with multiple flat bands.

The chiral spin liquid state improves on our parent state at the
mean-field level.  Since we will argue below that fluctuations stabilize
our state against the chiral mass term, let us here analyze why
the chiral state works at the mean-field level, and provide
a qualitative explanation of why fluctuations destroy the chiral
spin liquid.

The idea behind the chiral state results from the 
``Rokshar rules"\cite{rules}, which
argue that one should put flux $\frac{\pi}{2}$ through every triangle,
no flux through the hexagons, and have a total flux of $\pi$ through
the loop of length twelve that surrounds a hexagon and its six attached
triangles.  These rules are derived from considering individual
hexagons and triangles in isolation, and minimizing the mean-field
energy.

While our parent state appears to violate every one of the rules, except
the rule for the length-12 loop, the chiral spin liquid is in perfect
agreement with these rules!  Let us focus on one isolated triangle,
with $|t|=1$.  If there is no flux through the triangle, there are
two negative energy states with energy $-1$.  We can put two
particles in one state, and one in another, for a total energy of $-3$.
By adding $\frac{\pi}{2}$ flux, we have one state at energy $-\sqrt{3}$
and another state at energy zero.  By putting two particles in
the lowest energy state, we improve the energy of the system, and
introduce a chirality.

Now, consider the triangle coupled to the rest of the system.  If the
rest of the system strongly scatters the particles in the triangle, it
may no longer by appropriate to think of two particles in one state and
one in another.  One must instead think of each of the two negative
energy states of the triangle as each having average occupation
of one-and-a-half particles.  In that case, it is most advantageous to
put no flux through the triangle.

So, if the system to which the triangle is coupled is chiral, so that
all triangles in the system have the same flux $\frac{\pi}{2}$, then
the chiral spin liquid may work.  But if the triangle is coupled to
states which are not chiral, then the chiral spin liquid is destroyed.
One sees this even at the mean field level; a state in which triangles
have alternating flux $\pm\frac{\pi}{2}$ is significantly worse in
energy than our parent state.  Similarly, if one introduces sufficiently
strong dimerization $m_{12}$, one finds that the system is stable against
weak $m_c$.
Within the RG below, we will consider the fluctuations in the
masses $m_i$ and
show that they help stabilize the parent state against the chiral perturbation.

The spin solid of Marston and Zeng can also be obtained from the parent
state.  In this spin solid state, the idea is to look for dimer coverings
which maximize the number of ``perfect hexagons", hexagons on which
three of the bonds are covered by dimers.  Attached to these hexagons
are ``defect triangles", triangles on which no bonds are covered by
triangles.  Clearly, we wish to increase $|t_{ij}|$ on the perfect
hexagons, while decreasing it on the defect triangles.  This will
project onto the mass term $M_6$ on the 12-site cell that includes
the given perfect hexagon.

The perfect hexagons are then supposed to form a lattice.  We can
obtain this lattice by taking the triangular lattice of 12-site cells,
and placing perfect hexagons inside the 12-site cells on two out of the
three sublattices of the triangular lattice.  In this case the cells
containing perfect hexagons form a honeycomb lattice.  This gives
rise to a staggered mass state.  We
produce a mass term $M_6$ on two-thirds of the system, so that
the Dirac particles feel a net $M_6$ at zero momentum, as well as
a fluctuating $M_6$ at finite momentum.
\section{Naive Mean-Field and Projected Mean-Field}
To compare the energies of possible RVB states, including the various
mass perturbations of our parent state, we turn to the RVB mean field.
When we look for a mean-field solution of $t_{ij}$ in equation (\ref{mft1}),
it is known\cite{dimerize} that the best mean-field is a dimer covering.
Still, let us start by looking at results of the naive mean-field calculation,
and then later provide a projected mean field calculation.

Let us introduce the Green's function between sites, $G_{ij}$, defined to be 
the sum over all occupied fermionic states, $\psi$, of
\be
\psi^{\dagger}(i) \psi(j)
\ee
The fermionic energy is then equal to
\be
2 \sum_{<i,j>} G_{ij} t_{ji}
\ee
where the factor of 2 arises from the presence of up and down species
of fermion.

For our parent state, explicit calculation shows that, for
nearest neighbors $i,j$, 
\be
|G_{ij}|=0.221383...
\ee
From the mean-field condition for equation (\ref{mft1}), 
we find for the parent state that
$t=0.221383 J$, so by equation (\ref{vfeq})
\be
v_f=(0.0904...) J
\ee

We find that within the projected mean-field that the parent stable is
unstable to all of the massive fluctuations, including the chiral mass
fluctuation which will drive the system to a chiral spin liquid.  

For infinitesimal perturbations, the different 
nonchiral masses all provide an equivalent improvement in mean-field energy.  
To some extent this is due 
to the approximate low energy symmetry of the Dirac equation to rotating 
continuously among the different mass terms.  However, it is interesting
that lattice effects do not break this symmetry.  The reason is
the discrete lattice symmetry.  The change in energy for taking
$M=m_1 M_1+m_2 M_2 + m_3 M_3$ is, for small $m$, a quadratic form
in $m_i$.  Let this form be
\be
\label{qform}
\sum_{i,j} c_{ij} m_i m_j
\ee
The coefficients $c_{11},c_{22},c_{33}$ in this form must
all be the same due to lattice rotation symmetry.  Lattice translation
symmetry permits one to change the sign of any two of the $m_i$, and
prevents a nonvanishing $c_{ij}$ for $i\neq j$.  Therefore, for
small perturbations
the energy gain for introducing a mass must be dependent only on the
magnitude of the mass, not the particular mass term used.
For larger perturbations, the energy gains may depend on the particular
mass term used, and of all
the nonchiral mass terms, the system gains the most energy by
a mass $M_{12}$.

Now let us turn to the projected mean-field.
Instead of doing a full Gutzwiller projection, we will use an approximation
introduced by Hsu\cite{hsu}.  Within a variational Gutzwiller projection,
one minimizes the energy of the Hamiltonian (\ref{sham}).  Hsu's idea
at the lowest level of approximation is to note that the Hamiltonian is
a sum of terms $J \vec S_i \cdot \vec S_j$ over different neighbors $i,j$, and,
when evaluating the expectation value of each of these terms, to perform
the projection only on the given sites $i,j$.  At this level, the variational
principle corresponds to minimizing
\be
\label{hsmin}
\sum_{<i,j>} (\vec S_i \cdot \vec S_j)\approx
-\sum_{<i,j>} 6 \frac{|G_{ij}|^2}{1+16 |G_{ij}|^4}
\ee
over all possible $t_{ij}$, where $G_{ij}$ is determined by
$t_{ij}$.  

For our parent state, we find that 
$-6 \frac{|G_{ij}|^2}{1+16 |G_{ij}|^4}=-0.2832...$   By going to the
chiral spin liquid, the system improves the ground state energy by
roughly $2.9$ percent within the projected mean-field approximation.
By going to a state with staggered $\pm \frac{\pi}{2}$ flux through
each triangle, the systems worsens the ground state energy by
roughly $1.4$ percent.  Within this approximation the system is stable
against the nonchiral mass perturbations as all the nonchiral
mass perturbations worsen the ground state energy at this
level of approximation.  Again due to discrete lattice symmetry, 
the energy cost is independent of the particular mass term for small
mass, while for larger perturbations, the energy costs
differ, and the $M_{12}$ perturbation costs the least energy.
\section{Field Theory of Fluctuations}
Having discussed the naive and projected mean-fields for the problem,
we must include fluctuations about the mean-field.  To do this, we
will use an $SU(N)$ generalization of the original problem\cite{sun}, such that
the projection procedure of Hsu becomes exact.  After discussing how
to do this in the abstract, we will present the field theory for our
specific problem: it will have a number of interacting fields, including
fermions, gauge fields, the nonchiral mass terms discussed above, which we
will refer to as ``pion" fields, and the chiral mass term.

The approximation of Hsu ammounts to minimizing equation (\ref{hsmin}).
By introducing an auxiliary field
$\lambda_{ij}$ we can ``decouple" this sum of functions of
$G_{ij}$ and instead extremize the function
\be
\sum_{i,j} G_{ij} \lambda_{ji} + f(|\lambda_{ij}|^2)
\ee
where $f(|\lambda|^2)$ is
a Legendre transform of $\frac{|G|^2}{1+16|G|^4}$.

Then, we can interchange the order of extremizations, and 
extremize this quantity over $t_{ij}$ before extremizing over $\lambda_{ij}$.
We find that this is extremized at $t_{ij}=\lambda_{ij}$.  Then
we proceed to extremizing over the one remaining set of variables
$\lambda_{ij}$.  But since $t_{ij}=\lambda_{ij}$, we are
are equivalently trying to extremize the function
\be
\sum_{i,j} G_{ij} t_{ji} + f(|t_{ij}|^2)
\ee
over all $t_{ij}$.
Note, now, that $G_{ij} t_{ji}$ is exactly the kinetic energy of the
fermions.  So, finally, we are trying to
extremize
\be
\label{ton}
H=\sum_{<i,j>} \Bigl(\psi^{\dagger}_a(i) t_{ij} \psi_a(j)+h.c.\Bigr)+
\sum_{<i,j>}f(|t_{ij}|^2)
\ee

Returning to the language of functional integrals, we can introduce
an $SU(N)$ field theory for which the Hsu projection procedure becomes
exact.  We take a large $N$ limit in the number of fermion fields in
equation (\ref{ton}),
so that $a=1...N$.  Then, we integrate over all possible $t_{ij}$ in
that equation,
undoing the decoupling procedure above, and rewrite the
result in terms of spin operators.  We find
\be
\label{sunmft}
H=\sum_{<i,j>} \frac{\vec S_i\cdot \vec S_j}{1+16 \vec S_i\cdot \vec S_j}
\ee

We should note a few facts about this procedure.
When we demonstrate the equivalence of the large $N$ mean-field with the Hsu 
mean-field, it is the large $N$ limit that permits us to
ignore fluctuations in $t_{ij},\lambda_{ij}$, so that the decouplings
amounts exactly to taking a Legendre transforms; at finite $N$, the
decoupling of an interaction is not exactly a Legendre transform.
Further, we used the word ``extremize" above with
care: in some cases we maximize while in other cases we minimize, as in
some places the function $\frac{|G|^2}{1+16 |G|^4}$ has positive 
curvature while in other cases it has negative curvature.  This does
not provide any formal problems when performing the decoupling at
the level of functional integrals, so long as we correctly choose the
integration contour of $\lambda_{ij}$.  

The fractional operator,
$\sum_{<i,j>} \frac{\vec S_i\cdot \vec S_j}{1+16 \vec S_i\cdot \vec S_j}$ in
equation
(\ref{sunmft}) may be interpreted as a formal power series, so
that it includes operators of the form $(\vec S_i \cdot \vec S_j)^k$ for all 
$k$.  
At $N=2$,
this operator is equivalent to $\vec S_i \cdot \vec S_j$, up to a constant 
factor.

Finally, the above procedure is similar to the technique of introducing
biquadratic interactions, $(\vec S_i \cdot \vec S_j)^2$, 
into the Hamiltonian to
stabilize RVB states against dimerization.  We simply prefer the
above Hamiltonian as it reproduces exactly our desired mean-field theory.

Having defined a large $N$ theory with no fluctuations in the decoupling
fields, we next add in fluctuations.  Formally, this can be handled by
a $1/N$ expansion.  We will directly write the field theory at $N=1$ without
including explicit factors of $N$.  

The theory includes several modes.  There is the Dirac fermion field,
$\psi(x)$.  This is coupled to a fluctuating $U(1)$ gauge field,
$A^{\mu}(x)$, $\mu=t,x,y$.  By considering other fluctuations in
$t_{ij}$ we will 
also obtain a fluctuating chiral mass field that we will refer to
as $\sigma(x)$, and
a triplet of fluctuating nonchiral mass terms that we will group into
one ``pion" field $\pi^{a}(x)$, $a=1,2,3$.  If the pion field acquires
an expectation value, then the fermions will acquire a mass
$m_a=\langle \pi^{a}\rangle$.

For the field theory, we will suppress the velocity $v_f$.  At the level
of the bare action the
pi, sigma,  and gauge fields can have different velocities from the
fermi fields.  However, the greatest contribution to the effective action
of the bosonic fields arises from integrating over the relativistic
fermions, so that at low energies the velocity of the bosonic fields must 
be roughly equal to that of the fermionic fields.

The Lagrangian $L$ we will take is
\be
L=\int d^3 x\, L_f+L_A+L_M
\ee
where
\be
L_f= \overline \psi_{u,d}(x)
\Bigl(\gamma^{\mu} (A_{\mu}+i\partial_{\mu})+\gamma_0 M_a \pi^a +\gamma_0
M_c \sigma \Bigr) \psi_{u,d}(x)
\ee
\be
L_A=\frac{1}{4 \Lambda g_a^2}F_{\mu\nu}F^{\mu\nu}
\ee
\be
L_M=
\frac{1}{2 \Lambda g_{\sigma}^2}\sigma(x)\Bigl(\partial_{\mu}^2+m_{\sigma}^2
\Bigr) \sigma(x)+
\frac{1}{2 \Lambda g_{\pi}^2}\pi^a(x)\Bigl(\partial_{\mu}^2+(m^2_{\pi})^{ab}
\Bigr) \pi^b(x)
\ee
We have written the mass for the pion field as a matrix.
However, following the arguments for equation (\ref{qform}), the masses of
the different pion modes must be the same.  It is only after condensation
of the pion field, breaking the lattice symmetry, that the masses can
differ.

We have inserted factors of $\Lambda$, representing a lattice cutoff
scale, into the action to make the coupling constants dimensionless.
We have chosen to scale the bosonic fields so that all coupling
constant dependence appears in the action $L_A$ and $L_M$, not $L_f$.

In addition to the terms we have written, there 
must be a quartic interaction term for the $\pi$ and
$\sigma$ fields.  This term is necessary to stabilize the action if
the system spontaneously breaks a symmetry and has either $m_{\pi}<0$ or
$m_{\sigma}<0$.
Since we will be initially starting the renormalization
group of the next section with both such masses positive, we can temporarily
ignore the quartic term at high energy under the assumption that this
term is small.  If the system acquires an expectation value for
the $\pi$ fields, the quartic term will break the continuous symmetry down
to the lattice symmetry, and give a small mass $m_{\pi}^{\perp}$ for
the approximate Goldstone modes.  There will also be cubic terms
that give a mass to these modes.

Another interesting term we have left out is
$\overline\psi \gamma_{\mu}\gamma_{\nu}F^{\mu\nu}\psi$,
which can be added to change the $g$-factor of the Dirac fermions.  
In the absence of external fields,
there are two degenerate states of the Dirac equation at each energy.
For physical electrons, this reflects a spin degeneracy.  For the spinons
we consider, which already have a definite spin, this degeneracy 
instead reflects a {\it chirality} degeneracy, and we will refer to
it as such.  If the system has
an odd number of sites, and hence an unpaired spinon, not only
does the system have a net spinon, but it also has a net chirality,
which can be taken to be positive or negative.
Generically the $g$-factor will be non-zero.
\section{Renormalization Group}
We will consider the one-loop RG from the field theory.  We will see
that it is indeed possible for fluctuations to lead to a condensation of
the pion field.

For the gauge, pion, and sigma fields we will use a simple mode elimination
RG, with a cutoff $\Lambda$.  For the fermion fields, we will introduce
a set of massive regulator fields with masses of order $\Lambda$ and reduce the
regulator mass.  The choice
of the  particular mass terms for the regulator fields represents
a lattice breaking of the Goldstone symmetry.  It is possible to preserve the
needed lattice symmetry of equation (\ref{qform}) by introducing seven
regulator fields.  Four are ghost fields with masses proportional to
$M_1+M_2+M_3, M_1-M_2-M_3,-M_1+M_2-M_3,-M_1-M_2+M_3$ and the other three are
not ghosts and have masses proportional to $M_1,M_2,M_3$.

Initially the theory will have a cutoff $\Lambda_0$, defining the
lattice scale.  As we renormalize, we lower the cutoff $\Lambda$,
and rescale all distances and fields to keep $\Lambda$ fixed. 

We must take into account self-energy corrections to the fermions
from interactions with the bosonic fields, vertex corrections, and
self-energy corrections to the bosonic fields from vacuum polarization
bubbles.  If we were to take into account only the self-energy corrections
to the bosonic fields, and not the vertex and fermionic self-energy terms,
we would find that we are considering just the mean-field theory in
the bosonic fields.

We find the following RG equations:
\be
\frac{d{\rm ln}g_A}{d{\rm ln}(\Lambda_0/\Lambda)}=1-2\frac{g_A^2}{3}
\ee
\be
\frac{d{\rm ln}g_{\pi}}{d{\rm ln}(\Lambda_0/\Lambda)}=1+
\frac{3 g_A^2+g_{\pi}^2-g_{\sigma}^2}{2 \pi^2}
+{\rm vacuum \, polarization}
\ee
\be
\frac{d{\rm ln}g_{\sigma}}{d{\rm ln}(\Lambda_0/\Lambda)}=1
+\frac{3 g_A^2-3g_{\pi}^2-g_{\sigma}^2}{2 \pi^2}
+{\rm vacuum \, polarization}
\ee
\be
\frac{dm^{ab}_{\pi}}{d{\rm ln}(\Lambda_0/\Lambda)}=2+{\rm vacuum \, 
polarization}
\ee
\be
\frac{dm_{\sigma}}{d{\rm ln}(\Lambda_0/\Lambda)}=2+{\rm vacuum \, polarization}
\ee

We have avoided explicitly writing the vacuum polarization contributions
to the sigma and pion fields.  The vacuum polarization contribution to
the mass is regularization dependent, while the vacuum polarization
contribution to the coupling constant is ultraviolet convergent and
is dominated by the infrared contribution.

Fluctuations in the gauge field increase the coupling constants for
the pion and sigma fields.  This reflects the binding force due to the
gauge field between charged spinons, and the resulting tendency to break
chiral symmetry.
Further, we see that the coupling constant for the pion field
increases more rapidly than that for the sigma field, reflecting 
the destabilization of the chiral state by fluctuations in the pion field.

Thus, we see from the renormalization group that there is a range of
bare parameters such that the
theory will condense the pion field, producing a nonchiral mass term for the
fermions, even though at the mean-field level the theory would rather
produce a nonchiral mass term for the fermions.  

In the next section we will consider the low energy action after
condensation.  We will first discuss the mass term for the fermions that
appears.

Unfortunately, it is beyond
our ability to calculate the bare parameters in the field theory with
any precision, and so the mass of the fermion field is not something
we can compute.  
Let us instead take the mass of the fermion as an input from numerics, and
use that to check for consistency of our theory.  Extrapolating
finite size results from systems of up to 36 sites, one finds
that the system has a gap to triplet excitations which is of order
$J/20$ or less\cite{numerics1}.  
While the gap is decreasing even at the largest sizes,
it appears to be bounded below by roughly $J/40$.  Assuming that
the triplet excitations are made of pairs of spinons, we can estimate
the spinon gap as being half the triplet gap.  Further evidence for
the spinon gap being roughly half the triplet gap comes from odd-even
studies of the energy dependence on N\cite{numerics1}.  The fermion
mass is half the spinon gap, or one quarter the triplet gap.

Using this estimated spinon gap, and the calculated velocity of our
Dirac particles from RVB theory, we can obtain the correlation length
of the Dirac particles.  Taking the estimate of $J/20$ for the triplet gap, 
we find that the correlation length is roughly $8$ of our twelve-site unit
cells, large enough to include the $N=36$ numerical studies.

We can also estimate the strength of dimerization at the mean field
level, by asking how large a change in $t_{ij}$ is needed to produce 
the desired mass term.  Assuming that the dimerization is provided by a 
perturbation of the form $M_{12}$, one finds that the $t_{ij}$
12-site loops are increased by approximately $3.5$ percent, while the other
bonds are decreased by $3.5$ percent.
This is a relatively small amount of dimerization, and we expect that
only after a significant increase in system size will numerical studies
be able to detect this directly from a dimer-dimer correlation function.
\section{Low Energy Modes}
The remaining low energy modes after the pion field condenses 
are the Goldstone excitations of the pion field and the gauge excitations,
which we will argue provide the low energy singlet modes seen in numerics.
While numerical calculations have only probed systems up to $N=36$ sites,
which is relatively small considering that we take a unit cell of
12 sites, experiment also reveals a quadratic low temperature specific
heat.   This specific heat suggests that a bosonic mode with a linear
density of states survives to much larger scales, while the
insensitivity of the specific heat to weak magnetic fields suggests that
the mode is still a singlet.  In this section we will first
discuss the nature of the low energy modes and then the ultimate
fate of our pion and gauge excitations at large distances, including 
a gap for the pion from lattice effects as well as a confining phase for the
gauge fields.

Once the pion field condenses, the system is left with two pseudo-Goldstone
pion modes as well as gauge modes.
The gauge action is
\be
\label{gle}
L_A=\frac{1}{4 \Lambda g_A^2}F_{\mu\nu}F^{\mu\nu}
\ee
with 
\be
g_A^2\propto m
\ee
where $m=|\langle \pi \rangle|$ is the fermion mass.
The pion action will be a sigma model.  If we change the normalization
on the pion field so that $|\pi|=1$, we get the model
\be
\label{psg}
L_{M}=\frac{1}{2g^2}(\partial_{\mu} \pi^a)^2+
\frac{(m_{\pi}^{\perp})^2}{2g^2} \pi^1\pi^2\pi^3
\ee
where the coupling constant $g^2$ is proportional to $m^{-1}$ and
the mass $m_{\pi}^{\perp}$ represents the breaking of continuous symmetry
by lattice effects.

We have chosen the mass term for the pion to cause the pion to
prefer to condense in a way that gives rise to a
mass $M_{12}$.
In the projected mean-field calculation above, we considered states
invariant under the rotational symmetry, so that $|m_1|=|m_2|=|m_3|$.
This provided two inequivalent perturbations.  In one, we increased
$|t_{ij}|$ on hexagons and triangles; in the other we increased $|t_{ij}|$
on a loop of length twelve.  While at the mean-field level
symmetry breaking does not occur, from the projected mean-field calculation we
can still argue that the preferred symmetry breaking pattern would be
given by a mass matrix $M_{12}$.  Other patterns are of course possible,
and comparison with numerics provides some evidence
that a staggered mass is also a possibility, at least for small systems; 
in the conclusion we discuss
possibilities of numerically testing the preferred mass pattern.

In the continuum field theory, the pion mass
$(m_{\pi}^2)^{ab}$ is ultraviolet divergent.  However, lattice symmetry
forces the masses of the pion modes to be the same before condensation.
In order to use the continuum theory to estimate 
$m_{\pi}^{\perp}$ after condensation, we need to turn to
the cubic interaction terms in $\pi$.  
These are
\be
\int d^3 x\, g_3 \pi_1 \pi_2 \pi_3
\ee
with a cubic coupling constant $g_3$ that is generically of order unity.
Inserting an expectation value of $\pi$ of order $m$, we obtain a
quadratic term in $\pi$.  Including this quadratic term in $(m_{\pi}^2)^{ab}$,
this will cause the masses of the different pion modes to differ by order $m$
so that $m_{\pi}^{\perp}$ will be of order $m$.

However, we can obtain a better
estimate of the mass difference numerically from the
the projected mean-field
calculation of energy; while we did not obtain pion condensation at
the mean-field level, for a given expectation value of the pion
field, we can use the {\it difference} in mean-field energies
for various continuous rotations of the pion field to obtain an estimate of
the pion gap.  
Using the numerical estimate for the triplet gap, and hence the
estimate for the fermionic mass $m$, 
we have calculated the projected mean-field 
energies for taking $M=m M_{12}$ and $M=m M_{6}$.  The difference
in energies is $0.000396 J$ per 12-site cell, so that 
$m_{\pi}^{\perp}\approx \sqrt{0.000396 m J}$.
This is small enough that we can ignore this mass
for most purposes.  Evidently, the cubic
coupling constant $g_3$ is very small.

While the pion is already gapped by lattice effects, instantons
will gap the gauge field, leading to confinement of the
spinons.  
The gauge field describes compact QED in 2+1 dimensions, which is
confining for all $g_A$\cite{polyakov}.  
The gauge coupling is proportional to $m$,
so that the action for an instanton will be of order
\be
S \propto \frac{\Lambda_0}{m}
\ee
The instanton density is proportional to $e^{-S}$.
In the weak coupling limit, the instantons lead to a gap for the
gauge field of order $\Lambda_0 e^{-S}$.  As this is exponentially
small, we can ignore the gap in the gauge field. 
\section{Finite Size Systems and Chern Numbers}
In this section we will consider some effects of finite size systems to
begin comparison with numerics.  First we will consider some complications
in defining the parent state on systems with an odd number of sites,
which force the system to have some net flux.  Then we will show how this
leads to a degeneracy in the spectrum and nonvanishing Chern numbers for the 
states under twist in the
spin boundary conditions.  Finally, we will discuss some effects of
finite size for even size systems.

One of the most interesting results found in numerical studies of
odd size systems is a non-vanishing Chern number\cite{numerics1} for
the ground state of $\pm 1$.
This is a quantity that provides an analogue for a spin system of
the quantum Hall effect\cite{duncan}.  Since the Hamiltonian of equation
(\ref{sham}) does not explicitly break time-reversal symmetry, a non-vanishing
Chern number requires a spontaneous breaking of time-reversal symmetry.
However, the spontaneous breaking of time-reversal symmetry is not enough, as
other spin systems that break this symmetry have
vanishing Chern number\cite{duncan}; the \kl antiferromagnet
may be the first Hamiltonian with time-reversal and parity symmetry
where a non-vanishing Chern number has been observed.  In order to
understand the appearance of the Chern number, we must first understand
how to form the parent state on an odd size system.

The systems studied numerically have periodic boundary conditions, so that 
they live on a torus.  On system defined on a torus the net flux penetrating
the surface must be an integer multiple of $2\pi$.  One can also add
solenoid fluxes, $\theta_1,\theta_2$, defining the phase that the spinon 
acquires when traversing a topologically nontrivial closed loop around the
torus.  For simplicity, we will use coordinates on the torus which range from 
0 to $2\pi$ in both directions, although in actuality for the \kl
lattice the systems considered are not square.

The parent state has $\pi$ flux through each hexagon.  On a
system with $N$ sites, there are $N/3$ hexagons, and so on a system with
an odd $N$, one would like to have a net flux through the system that
is an odd multiple of $\pi$.  This is not possible, and so the system
must have some additional flux so that the total is a multiple of $2\pi$.
For example, on a system with $N=27$, there are 9 hexagons, so the
system can put $\pi\pm\frac{\pi}{9}$ flux through each so that the
total flux is either $8\pi$ or $10\pi$.  The system then must become
chiral and break time reversal symmetry since it cannot construct
the parent state.

A qualitative way of describing this effect is to say that
for a system with an odd number of sites, there is an unpaired
spinon, which has a chirality.  The spinon then couples to the gauge
field and produces a flux.

Given this net flux through the system, let us consider the effective
Dirac equation for the spinons.  The results we get for the
Chern number do not rely on the Dirac description, and can be derived
directly from the lattice model; we feel that the Dirac method is more
elegant and gives more physical insight.

The Dirac particles feel 
the extra flux that has been added, and so the spinons move in a magnetic
field, such that the net flux the spinons feel is exactly $\pm \pi$.
Again, there seems to be a contradiction, since it is not possible
for the system to have a net flux of $\pm\pi$ flux through the torus.
The answer to the contradiction is that the Dirac particles have
an extra chirality index.  So, in addition to including solenoid fluxes
for the Dirac equation, the Dirac particles can change chirality when
completing a loop around the system.  Let us then generalize the
solenoid flux to a pair of 4-by-4 matrices $U_1$, $U_2$, describing the
change in the wavefunction when the particle completes a loop.

Then, when the particle traverses a loop around the torus from
$(0,0)$ to $(0,2\pi)$ to $(2\pi,2\pi)$ to $(2\pi,0)$ to $(0,0)$, 
the wavefunction gets multiplied by 
\be
\label{ab}
-U_1 U_2 U_1^{\dagger} U_2^{\dagger}
\ee
where the minus sign is from the magnetic flux through the torus.
Since equation (\ref{ab}) must be equal to 1, we find that
$U_1$,$U_2$ necessarily anti-commute.

One may regard the matrices $U_1,U_2$ as arising from a non-Abelian
gauge field connecting opposite chiralities of the spinons.  The
commutator of the matrices represents an additional flux of $\pi$ from
the non-Abelian field, giving a total flux of $2\pi$ on the torus.  
The extra $\pi$ flux from the non-Abelian field is the flux that arises
from having an odd number of hexagons on the lattice, so that when the
particle completes the given loop around the lattice it has enclosed an odd
number of $\pi$ fluxes.  One sees that the non-Abelian flux is localized
at a point, although one must be careful that this localization at a point 
does not imply a breaking of translational symmetry.  

The addition of matrices $U_1,U_2$ is natural from the lattice point
of view.  The unit cell which includes both chiralities of Dirac particles
is 12 sites, while the smallest unit cell possible for the parent
state is 6 sites.  
Since there is no way to cover an odd size lattice with 6 or 12 site
unit cells, something must scatter between chiralities, as when
the particle completes a loop it has changed between chiralities.

To give a very simple example of this, consider a one-dimensional
ring with an odd number of sites.  The natural unit cell for the
a one-dimensional chain is two-sites, to include both Dirac points.
If the particle moves around an odd-length ring, two sites at a time, it
must return to the starting point displaced by half a unit cell.

To give a slightly more complicated example, consider the $\pi$-flux
phase of the square lattice\cite{cav} for a system of 9 sites, shown in figure
5.  Solid lines represent bonds within the cell of 9 sites, while dotted
lines represent bonds to provide toroidal boundary conditions.  There
are 9 squares in the system, and so there will be $\pi\pm\frac{\pi}{9}$
flux through each square.  4 of the squares lie within the cell and
are labeled 1-4, another 4 lie to the sides and are labeled 5-8,
while the 9th square lies in the corner.  The natural unit cell for
the Dirac particles is 4 sites, so when a particle completes a loop
around the torus it is displaced by half a unit cell.  The
wavefunction is multiplied by a matrix $U_1$ for a loop in
the $\hat x$-direction and a matrix $U_2$ for a loop in the $\hat y$-direction.
Precisely due to the odd number of squares, one finds 
that the matrices $U_1,U_2$ anti-commute.
It is natural to think of
the non-Abelian flux as arising from the $\pi$-flux through the 9th square, 
on the corners.

Returning to the \kl lattice,
let us now look at the wavefunctions of the Dirac equation with this
magnetic flux.  It is convenient to find the wavefunctions by enlarging
the torus by a factor of two in each direction, as shown in figure 5.  
The $+$ and $-$ symbols in the figure denote the chirality of the particle
in each quadrant.  When the particle completes a circuit on the original
torus, it changes chirality, and hence moves into a different quadrant
of the enlarged torus, while picking up a phase.  

The net flux on the enlarged torus is equal to 4 times the
flux on the original torus, or $4\pi$.  One might imagine that there
will be extra sources of $\pi$-flux on the enlarged torus at the
points where the quadrants meet.  However, since the $\pi$ non-Abelian flux
is purely a result of an odd number of hexagons on the original
torus, we can drop the extra sources of flux on the enlarged torus,
and we are left with an explicitly translationally invariant problem
of a Dirac particle moving in a constant magnetic field.

On the enlarged torus, the Dirac equation becomes a two-component
equation
\be
\Bigl(i E \sigma_z +  \sigma_x (A_x+ i\partial_x) +\sigma_y 
(A_y +i\partial_y)\Bigr) \psi=0
\ee
Taking the square we find
\be
\label{landaul}
E^2 \psi=
\Bigl((i \partial_x-A_x)^2 +
(i \partial_y-A_y)^2 +\sigma_z B\Bigr)\psi
\ee
This is the well known equation for a Dirac particle in a magnetic
field, and has Landau levels.

Since the net flux through the enlarged torus is 
equal to $4\pi$, there are two-states in each Landau level for
each $\sigma_z$, hence four states for each Landau level in total.   
As we are dealing with a two component equation,
only one sign of $E$ is allowed in equation (\ref{landaul})
for given $\sigma_z$.
 
Therefore, the energy levels on the enlarged torus are doubly degenerate.
However, the enlarged torus has an unphysical degree of freedom: opposite
quadrants describe the state of the particle on the original torus.
So, the energy levels on the original torus are only singly degenerate 
and the spectrum is discrete with one level at zero energy.  This is
the relativistic generalization of Landau levels.  For
an odd size system, all Landau levels below the zero energy are occupied,
and hence filled, for
both spin up and spin down particles, while the zero energy level
is occupied only by one unpaired spinon.  

Numerically, the ground state of the many-body system 
has been seen to have an extra degeneracy factor of two, beyond the trivial
spin degeneracy.  This is a consequence of the spontaneous generation of the 
magnetic field, so that the system can pick either sign for the field.

An interesting way of viewing the Landau levels is that we have $\pi$ Abelian 
flux, implying that there
are $1/2$ states per Landau level.  Multiplying the $1/2$ by a factor of
two for chirality degeneracy, we get one state per Landau level.

We can now introduce the Chern number of the system, which characterizes
a transverse response of spin currents.  Let us adjust the
boundary conditions of the system so that
\be
\label{bc1}
S^{\pm}(x,y)=e^{\pm i \phi_1} S^{\pm}(x+2\pi,y))
\ee
\be
\label{bc2}
S^{\pm}(x,y)=e^{\pm i \phi_2} S^{\pm}(x,y+2\pi)
\ee
where $\phi_1,\phi_2$ are angles.

If the ground state is a wavefunction $\Psi$, then the Chern number is
defined as the integral
\be
\label{chern}
\frac{1}{2 \pi}\int \int \langle 
\frac{\partial\Psi}{\partial\phi_1}|
\frac{\partial\Psi}{\partial\phi_2}\rangle \, d\phi_1 \, d\phi_2
\ee
This number is quantized, and non-vanishing only for complex states.  Since
the Hamiltonian does not break time-reversal symmetry, 
complex conjugate states are degenerate with opposite Chern numbers.

In the presence of these boundary conditions, the spinon boundary conditions,
with additional self-generated fluxes $\theta_1,\theta_2$, become
\be
\label{sbc1}
\psi^{\dagger}_{u}(x,y)=e^{i \frac{\rho^u_1}{2}
} \psi^{\dagger}_{u}
(x+2\pi,y)
\ee
\be
\label{sbc2}
\psi^{\dagger}_{d}(x,y)=e^{i \frac{\rho^d_1}{2}}
 \psi^{\dagger}_{d}
(x+2\pi,y)
\ee
where
\be
\rho^{u,d}=(2 \theta_1 \pm \phi_1)
\ee
and similarly for the other direction.  The $2\pi$ periodicity in $\phi$
is not obvious from equation (\ref{sbc1},\ref{sbc2}), but the ability
of the system to adjust $\theta$ produces the desired periodicity.

To give a simple example of how a system can adjust $\theta$, 
consider a system of 4 sites on a ring.  In the absence of a twist in boundary
conditions $\phi$, the system places $\theta=\pi$ flux through the ring.  As
$\phi$ increases, $\theta$ remains equal to $\pi$, and the mean-field
energy of the system gradually increases until $\phi=\pi$.  At this
point, $\theta$ jumps to $0$, and the mean-field energy begins to
decrease for increasing $\phi$.  
So, as $\phi$ varies from $0$ to $2 \pi$ and $\theta$ jumps as described above,
we find that $\frac{\rho}{2}=\frac{(2 \theta+\phi)}{2}$ 
varies from $\pi$ to $\frac{3\pi}{2}$ to $\frac{\pi}{2}$ to $\pi$.
In order for $\theta$ to jump like this, the spinon states with 
$\frac{\rho}{2}=\pi \pm \frac{\pi}{2}$ {\it must be degenerate}.

Now, we can look at the Chern number of the system, assuming non-interacting
spinons.   It then amounts to a Chern number calculation of the fermionic
wavefunctions.  Assuming non-interacting spinons, we can 
get the change in $\Psi$ in equation (\ref{chern}) from the change in
the spinon wavefunctions.

While in general the wavefunction gets multiplied by a matrix on
moving around the original torus, only the $U(1)$ part of this matrix adjusts
in response to changes in $\phi$.  The $U(1)$ part of the matrix is just
the angle $\theta$.  Carrying out the calculation on the enlarged torus,
we find that the boundary conditions become
\be
\label{sbce}
\psi^{\dagger}_{u}(x,y)=e^{i \rho^u_1 } \psi^{\dagger}_{u}
(x+4\pi,y)
\ee
and similarly for down spinons.  

So, we wish to compute
\be
\label{spchn}
\frac{1}{2 \pi}\int \int \langle 
\frac{\partial\psi}{\partial\rho^{u,d}_1)}|
\frac{\partial\psi}{\partial\rho^{u,d}_2)}
\rangle 
\frac{\partial\rho^{u,d}_1} {\partial\phi_1}
\frac{\partial\rho^{u,d}_2} {\partial\phi_2}
\, d\phi_1 \, d\phi_2
\ee
summed over all spinon wavefunctions $\psi$.

In equation (\ref{sbce}) the periodicity in $\phi$ seems obvious even without
$\theta$, as on the enlarged torus the periodicity of the spinon
wavefunctions in response to a twist in boundary conditions is halved in both
directions.  
However, we have introduced the degeneracy of two on the 
enlarged torus representing the fact that on the original torus the 
wavefunctions are periodic in 
$(\rho_1,\rho_2)$ with
periods $(0,2\pi)$ and $(\pi,\pi)$ but not with period $(0,\pi)$,
and as a result only half the possible wavefunctions on the enlarged
torus are physical.

In order to keep the wavefunction in the physical sector,
$\theta$ must jump discontinuously by $\pi$ as $\phi$ changes,
and as a result for
a given spinon state we only integrate equation (\ref{spchn})
over half the torus of possible phases $(\rho_1,\rho_2)$.
The fact of integrating over half the torus, or equivalently
the fact that the Landau levels contain one physical and one unphysical
wavefunction, does not prevent a defining of the Chern number for
the spinon wavefunctions.  When $\theta$ jumps it connects 
two degenerate states, and so the contribution of equation (\ref{spchn})
to equation (\ref{chern}) must still be quantized as an integer which
we can still refer to as a Chern number for the spinon.

Within the lattice formulation there are no further conceptual problems
and we must simply compute the integrals, but
within the continuum Dirac equation we must account for the negative
energy sea.  The correct understanding of this was found by 
Haldane\cite{duncan2}.

One must add a massive regulator field, and compare the
difference in Chern
number between the massless and massive fields.
The massive field has the same Landau level spectrum, but no zero mode.
So, the difference in Chern numbers is due to the zero mode, which
sits in the lowest Landau level.  There are two states in the
lowest Landau level, one physical and one unphysical.  It is
the physical state that carries the Chern number 
of $\pm 1$, giving the ground state of the
spin system a net Chern number of $\pm 1$, as seen numerically\cite{numerics1}.

We expect that the low-lying states will continue to have an odd
Chern number, in agreement with numerical results.
If a particle-hole pair is excited within the
Dirac band near the Dirac point, the Chern number will not change.
If the particle is excited from the band edge, the Chern number can change 
by $\pm 2$.  Only if
a particle is excited from the flat band to the Dirac band can the Chern
number change by $\pm 1$, giving rise to an even Chern number.
However, these states will be much higher in energy.  One can also
consider excited states with
net Abelian flux be equal to $3\pi,5\pi,...$ 

It is very interesting to think about these possible excited states 
which may have more than $\pi$ flux for the Dirac particles. 
The two-component particles carry a $\sigma_z$ index, which will couple to
the magnetic field.  If a large field is induced, a
number of spinons of the same $\sigma_z$ will be produced in the zero mode,
so that the total number of spinons in the zero mode is odd. 
For one spinon we had one filled Landau
level, with one particle.  With several spinons one might be able to construct
fractional Hall states of spinons.

We have argued that in the thermodynamic limit the system will
acquire a mass.  On an odd size lattice, the mass term must change 
sign somewhere, as the lattice cannot be tiled with 12-site unit cells.
At the domain wall where the mass changes sign, one expects to
trap a midgap state, so there still should be a zero mode, even
with mass.  This may permit the nonvanishing Chern number to survive.

Returning to even system size,
let us consider the size dependence of the triplet gap.
The energy of the spinon is
$E=\sqrt{(v_f k)^2+m^2}$.  In the absence of a solenoid flux, the
smallest $k$ would be equal to zero, but
by creating a solenoid flux the energy can be improved and
the smallest $k$ will be of order the inverse linear dimension of
the system, or $N^{-1/2}$.  As a result, the triplet gap is
decreasing with system size, in agreement with numerics.
By twisting the spin boundary conditions one may be able to reduce the gap
to $S_z=\pm 1$ excitations.
It would be interesting to look for this effect.

Further, in the presence of these solenoid fluxes, other fermionic states
with $k \neq 0$ will become approximately degenerate with the $k=0$
fermionic state.  This means that a spatially varying mass term which
scatters between $k$ states can open a gap just as well as the spatially
constant mass term can.  This will be important when we consider the
low energy Goldstone excitations on finite size systems, below.
\section{Goldstone Modes, Tower States, and Numerics}
After breaking a symmetry, and giving mass to the fermions, the
system is left with low energy pion and gauge modes.
Above, we argued that the gap for these
modes is too small to be seen in numerical calculations.  In this section, we
will treat these modes as gapless and discuss the energy spectrum that
results for finite size systems to compare with numerical
calculations.  

It is known\cite{tower} that breaking a continuous symmetry gives
rise to two kinds of low energy modes.  First, there are the Goldstone modes
with non-zero wavevector $k$.  In the case of our pion and gauge 
modes, the energy
is then proportional to $k$.  For a $2+1$ dimensional system with
$N$ sites, the lowest $k$ is proportional to $N^{-1/2}$ and so
the lowest Goldstone excitation has energy proportional to
$N^{-1/2}$.

Second, there is the ``tower" of $k=0$ modes.  These correspond to
global rotations of the entire system, and have an energy proportional
to $N^{-1}$.  

Numerical diagonalization\cite{numerics3}
of the {\it triangular} lattice
Heisenberg anti-ferromagnet, which has N\'eel order, shows very clearly the
distinction between the tower of states, and the $k\neq 0$ states (spin
waves).  However, no such distinction is found in the \kl 
lattice\cite{numerics4}, no
separation between low energy modes of energy $N^{-1}$ and $N^{-1/2}$.
Within our model, this is to be expected for $N=36$.  Since the effective 
action of the pion and gauge fields arises from integrating out the fermions,
this action must be approximately relativistic, with the same velocity
as the fermions.  So, even without explicit calculation, we can obtain
the energy of the lowest $k\neq 0$ mode directly from the velocity
appearing in Dirac equation.  For the largest numerical diagonalizations,
systems with $N=36$ total sites or 3 of our 12-site unit cells, this energy 
turns out to be of order the triplet gap, and so this Goldstone mode is too 
high in energy to appear in the continuum of low energy singlets.

The only states that will be observed in numerics are states in the
tower.  We can obtain the energy of these states from 
equation (\ref{psg}), assuming that $\pi(x)$ is constant.  Then we
get, assuming small $m_{\pi}^{\perp}$,
\be
\label{psg2}
L=\frac{N\Lambda_0^{-2}}{g^2}(\partial_{t} \pi^a)^2
\ee
where $N\Lambda_0^{-2}$ is the area of the system.  
The states of this will be spherical
harmonics, perturbed by mass term.  With $g^2\propto m^{-1}$, these states
will have an energy proportional to 
\be
\frac{\Lambda_0^2}{Nm}
\ee
and so
for sufficiently large $N$  will be below the triplet gap.  A more precise
knowledge of the prefactors will be needed to determine whether this
is low enough to correspond to the low energy modes seen numerically.

There will also be ``tower" states for the gauge field, which correspond
to different solenoid fluxes through the system.  The energy for
these can be obtained from equation (\ref{gle}) assuming that $A_{\mu}$
is constant over the sample.  To get the energy for these we have to
remember that the gauge group is compact, and realize that
$F^{\mu\nu}$ is derived from a set of $U(1)$ matrices with a lattice
length $\Lambda_0$.  Then, the energy of the gauge states is
proportional to
\be
\frac{m}{N}
\ee
which is definitely below the triplet gap and certainly small
enough to be the origin of some of the low energy modes in numerics.

There is one puzzle involved in the tower states.  It was observed numerically
that, on 36 site samples, the energy of the lowest energy state of
the system at given total momentum did not vary appreciably across
the Brillouin zone\cite{numerics1}.  
This may seem to be in contradiction to the
hypothesis that the low energy states come from the tower.  The resolution
of this may lie in realizing that there are only 4 inequivalent points
in the Brillouin zone of the system.  One of these is at $k=0$, while
including a constant nonvanishing mass term $M_{12}$ also breaks translational
symmetry on the \kl lattice, and can give one more point in the
Brillouin zone.  To obtain the last points in the Brillouin zone, one
must remember that for small systems there can be other symmetry breaking
patterns with staggered mass, 
breaking translational symmetry in different ways, as discussed
in the section on finite size effects.  

This would imply that for sufficiently large system sizes
one would find a more significant variation in the energies across
the Brillouin zone, as only some of the states could be obtained from the 
tower.  One would also find for 36 site systems
that the spinon solenoid fluxes would change
under a twist in spin boundary conditions, and so the fermionic states
at different $k$ would lose their degeneracy, leading to a change in
energy of some of the $k \neq 0$ states when varying boundary conditions.

Another possibility is that, even
for infinite systems, short-distance effects lead to the production of
a staggered-mass pattern, in the style of the ``perfect hexagon" state\cite{mz}
discussed above.
The staggered mass pattern gives rises to an enlarged unit cell, and makes
it possible to get low energy states at the other points in the
Brillouin zone.
\section{Further Comparison With Numerics}
In addition to the existence of the low energy states, we make
further comparison with numerics for the many-body density
of states and the dimer-dimer correlation functions.

Assuming the existence of a low-energy bosonic mode with linear 
dispersion relation, so that the single-particle density of states
scales linearly with energy, one would expect the many-body density of states 
at energy $E$ to scale for large systems
as an exponential of $E^{2/3}$.  The quadratic behavior
of experimentally measured specific heat is in agreement with this.

Since we have argued
that the low-energy states in numerical calculations are largely
``tower" states, it is impossible to extract the density of states in
a large system from the finite size density of states.  The true exponential
growth can only be seen when the $k \neq 0$ modes become important.

In this regard, the numerically measured\cite{numerics1} {\it quadratic
many-body density of states} at very low energies does not
say anything about the dispersion of the Goldstone modes.  Instead, it is
a reflection of the fact that if a finite number of ``tower" modes 
are excited then the many-body density of states is a power law.

For the dimer-dimer calculations we can make a more direct comparison with
a numerical calculation of these correlations on a 36 site 
system\cite{numerics2}.  In the thermodynamic limit, the dimer-dimer
correlation function should be long-ranged, reflecting the existence of
a non-zero $m_{12}$.  

However, there is also a short-range fermionic contribution to
the dimer-dimer correlation functions, and 
for 36 site systems, so that the system size is smaller than the correlation
area of the fermions, we can ignore the effect of a non-zero $m_{12}$ on
the dimer-dimer correlation function and directly study the correlation
functions of massless fermions.

If we are interested in a dimer-dimer correlation function
\be
C_{(i,j)(k,l)}=
\langle (\vec S_i \cdot \vec S_j) (\vec S_k \cdot \vec S_l)\rangle-
\langle (\vec S_i \cdot \vec S_j)\rangle \langle (\vec S_k \cdot \vec 
S_l)\rangle
\ee
we can express this, under the assumption of weak spinon interaction, directly
in terms of the spinon Green's functions.  Writing each spin operator in
terms of spinons and considering various contractions we obtain
\be
\label{ddf}
C_{(i,j)(k,l)}=
-4 {\rm Re} (G_{ij} G_{jk} G_{kl} G_{li})-
4 {\rm Re} (G_{ij} G_{jl} G_{lk} G_{ki})-
{\rm Re} (G_{ik} G_{kj} G_{jl} G_{li})+
\frac{1}{2} |G_{ik}|^2 |G_{jl}|^2+
\frac{1}{2} |G_{ij}|^2 |G_{jl}|^2
\ee
To obtain quantitatively accurate answers, we must include the effects
of projection within an approximation like that used above, projecting on
site $i,j,k,l$, requiring that there be one fermion on each of these four sites.
We have done this at 3 different levels of approximation.

 At  the lowest level, we have noted
that the wavefunction before projection will have one fermion on
each of these sites is roughly the product of the probability that it
will have one fermion on each of $i,j$ by the probability that
it will have one fermion on each of $k,l$.  Given that
$|G_{ij}|=0.221383...$ for neighboring $i,j$, we should replace equation
(\ref{ddf}) by
\be
\label{ddf2}
C_{(i,j)(k,l)}=
\kappa \Bigl(-4 {\rm Re} (G_{ij} G_{jk} G_{kl} G_{li})-
4 {\rm Re} (G_{ij} G_{jl} G_{lk} G_{ki})-
{\rm Re} (G_{ik} G_{kj} G_{jl} G_{li})+
\frac{1}{2} |G_{ik}|^2 |G_{jl}|^2+
\frac{1}{2} |G_{ij}|^2 |G_{jl}|^2\Bigr)
\ee
where
\be
\kappa=(1.9264)^2
\ee
is the desired factor of $(\frac{4}{1+16|G|^4})^2$.

At a more refined level, we have carried out the computation projecting
on all four sites exactly.  At the third level of approximation we
have started to project out onto additional sites as well.
We have calculated the correlation functions in this approximation, in
the limit of an infinite system size, for pairs of bonds that can both be 
written in the same 12-site unit cell.  We compare the result to
results from numerics on a 36-site system\cite{numerics2}.  We do not
consider pairs of bonds that cannot be written in the same unit cell,
as at this separation, finite-size effects will become important in the
numerics and the comparison will become impossible.  We could in principle
improve on our comparison with numerics by computing the spinon correlation
functions in a 36-site system also, in which case it should be possible
to compare all bonds, but we have not done this.

The results are shown in table I, where in the column
Theory I we show the
simplest approximation, and in the column Theory II 
we show the second level of approximation.  
Qualitatively, the theory works quite
well on the signs even at this level, getting 9 out of 12 correct.  
The only signs that
the theory gets wrong at this level occur when the theory predicts a very small
value ($<.01$) for the correlation function.  

To improve
this result, we included the third level of approximation in which
we also project out onto a fifth site $(m)$ for those dimer correlation 
functions
such that there is one and only one site $(m)$ which neighbors both
dimers.  In the column Theory III we show this level of approximation,
as well as the particular site $(m)$ that we picked.  Once this
is done all the signs work out for 11 out of 12 correlations,
and the qualitative agreement is almost perfect.

The magnitudes work out
less well, as most of the dimer-dimer correlations are far too small within
the spinon calculation.  However, RVB calculations are quite poor at
getting long range correlations without including some gauge fluctuations.
For example, in the one-dimensional Heisenberg antiferromagnet, the
spins on the same sublattice are uncorrelated\cite{path}.
By including gauge fluctuations, this result can be substantially
improved\cite{mudry2,patrick}.  

For the largest positive correlation
function, $C_{(6,7)(1,8)}$, the magnitude does work out 
well even at the simplest approximation.  However, 
for the largest negative function, $C_{(6,7)(11,5)}$, the
magnitude is off by roughly a factor of 5, until we go to the third level
of approximation, at which point the magnitude becomes roughly correct.
We can hope that a better
inclusion of fluctuations will improve these results, just as it
has done for the one-dimensional chain.  Perhaps projecting on an
entire 12-site cell would give better results, as suggested by the
improvement in the results in column III.
Detailed calculations of projection for some trial wavefunctions
on the \kl lattice have been performed by Hsu and Schofield\cite{hsus}.
A similarly detailed calculation for our parent state would
be of interest.
\section{Conclusion}
In conclusion, we have constructed an RVB state on the \kl lattice,
the ``parent state", which has a Dirac structure.  Consideration of the various
mass perturbations to the Dirac equation unifies several other previously
suggested long-range states of the \kl lattice Heisenberg antiferromagnet,
with the exception of the BCS\cite{hsus} state and the bosonic $Sp(2N)$ 
state\cite{subir}.

While at the projected mean-field level the chiral spin liquid appears to be
the best RVB state, we have argued by a renormalization group treatment
that fluctuations provide a mechanism for stabilizing a state with a
nonchiral mass term.  The numerical evidence also argues against a
chiral state.  We have then proceeded to explicit comparison with numerics,
taking as input only one quantity, the triplet mass gap.

The physical idea behind our construction is that, given massless fermions,
the system must ultimately try to break some symmetry to give mass to the
fermions.  There are many ways of doing this, but they all correspond to
either introducing chirality, or to introducing some kind of spin-solid state
in which the system dimerizes the $t_{ij}$.  If we ignore the chiral
state, we must have some kind of spin solid: we have proposed one possible
spin solid and its attendant pseudo-Goldstone excitations.  Other
spin solids are possible, and may in fact be realized in favor of our
proposal, but the general principle that the massless fermions must
break a symmetry and produce pseudo-Goldstone excitations should be robust.
Two other possibilities for spin solids that must be considered are the
constant $M_6$ solid, and the perfect hexagon system with spatially varying
$M_6$.

Experimentally, it may be possible in principle to detect the
spin-Peierls order.  Although this as appears as long-range order only in
4 spin correlation functions, it should give rise to a short-range
oscillatory piece in the spin-spin correlation function.  Since
the spin-spin correlation function decays exponentially and
the dimerization is weak, this would be very difficult to detect, but
in principle it is possible via neutron scattering. 

Theoretically, more work is needed on the fluctuations about the parent
state.  Doing a Gutzwiller projection of the wavefunctions on a 36-site
lattice should enable much more direct comparison with numerics, especially
for the dimer-dimer correlation functions.

Numerically, it may be possible to confirm the identification of
the low-energy states with tower states from symmetry breaking.  If
one computes a dimer-dimer correlation function with the same pair of
sites
taken at two different times it may be possible to see the oscillations
in the mass field.  This may be difficult, though, given the relatively
weak amount of dimerization present and the problem of extracting the
contribution of the mass term to dimer-dimer correlation functions from
the background of the fermionic contribution.  Similarly, one
can try to compute a susceptibility to spin-Peierls ordering
by explicitly dimerizing coupling constants $J$ in equation (\ref{sham}).
If indeed the system wishes to spontaneously order in the thermodynamic
limit, then the susceptibility to dimerization should be large.  This
may make it possible to unambiguously determine whether the system
prefers the $M_{12}$ or other ordering pattern.

It might also be interesting numerically to look at model systems in which
the Hamiltonian has additional terms coupling to the chirality operator
on each triangle.  This might make it possible to probe the stability
of the system to the chiral mass term, as well as providing some interesting
states in which the Dirac fermions are moving in a large net magnetic field.
\section{Appendix: An Interesting Flat Band Case}
When the system has flux $\pi/4$ through each triangle
and flux $\pi/2$ through each hexagon the band structure becomes very
peculiar.  Using a 12-site unit cell, we find that the lowest band
is doubly degenerate, and almost exactly flat.  The next band above
that is quadruply degenerate and exactly flat.  The higher bands, which
are all empty, are not flat.

It is very unlikely that any such state could be stabilized.  We have argued
above the the \kl lattice is not a chiral spin liquid.  However,
it may be possible to add chirality operators to the Hamiltonian to
tune the flux through the triangles.

In this case, the physics of the flat band state would be very amusing.
We are used to the fact that in, for example, the one-dimensional
Heisenberg antiferromagnet, one can deduce that the spin-1 excitations
are composite objects of two spinons by looking at the excitation spectrum.
Since the energy of the spin-1 object is a sum of two different energies,
there is a continuum of possible energies.  When, however, one of the two 
spinons is a hole excitation from a flat band, the energy is constant over the
band, and there is no sign of the composite nature of the spin waves
when looking at their energy spectrum.  One would have a situation with
one spinon hopping freely over the lattice, while the other spinon
sits unmoving on a given unit cell!
\section{Acknowledgements}
I would like to thank S. Sondhi for suggesting the problem of the
\kl antiferromagnet, and for many useful discussions and insights.
I would also like to thank R. Moessner for useful discussions
on theory and experiment in frustrated antiferromagnets, V. N. Muthukumar
for discussions on RVB ideas, A. Vishwanath for discussions on
Chern numbers, and S. Sachdev for clarifying the estimate of
the pion gap.

\begin{table}
\label{table1}
\caption{Comparison of $C_{(i,j)(k,l)}$ between numerical and spinon 
calculations.  See figure 1 for labeling of points in unit cell.
See text for discussion of various approximations.}
\begin{tabbing}
\hskip10mm \= (i,j)(k,l) \hskip 10mm \=  Numerics  \hskip 10mm \= Theory I
\hskip10mm \= Theory II \hskip10mm \= Theory III \hskip10mm \= $(m)$ \\
\> (6,7)(1,8) \> 0.04337 \> 0.08242\> 0.06249\\
\> (6,7)(8,2) \> -0.01416 \> -0.01107\> -0.001318\> -0.01686\> (1)\\
\> (6,7)(9,3) \> -0.00646 \> -0.001041\> -0.0004279\> -0.022153\> (8)\\
\> (6,7)(10,3) \> 0.01178 \> 0.001043\>0.0005004\\
\> (6,7)(10,4) \> -0.01045 \> -0.001040\> -0.0005008\\
\> (6,7)(11,5) \> -0.06510 \> -0.02499\> -0.0112132\>-0.083829\> (12)\\
\> (11,10)(7,8) \> 0.01221 \> 0.018885\> 0.00938\\
\> (11,10)(9,8) \> -0.00113 \> -0.01824\> -0.0025\> -0.01395\> (3) \\
\> (11,10)(2,9) \> 0.00108 \> 0.0001619\> 0.00559\> -0.007171 \> (3) \\
\> (6,7)(11,4) \>  0.01322 \> -0.0009778\> -0.000327 \> 0.001416 \> (12) \\
\> (11,10)(2,8) \> 0.00045 \> -0.004377\> -0.001964 \> 0.00386 \> (9) \\
\> (6,7)(9,2) \> 0.01322 \> -0.0009788\> -0.000328 \> 0.001415 \>(8) \\
	\end{tabbing}
	\end{table}
	\begin{figure}[!t]
	\begin{center}
	\leavevmode
	\epsfig{figure=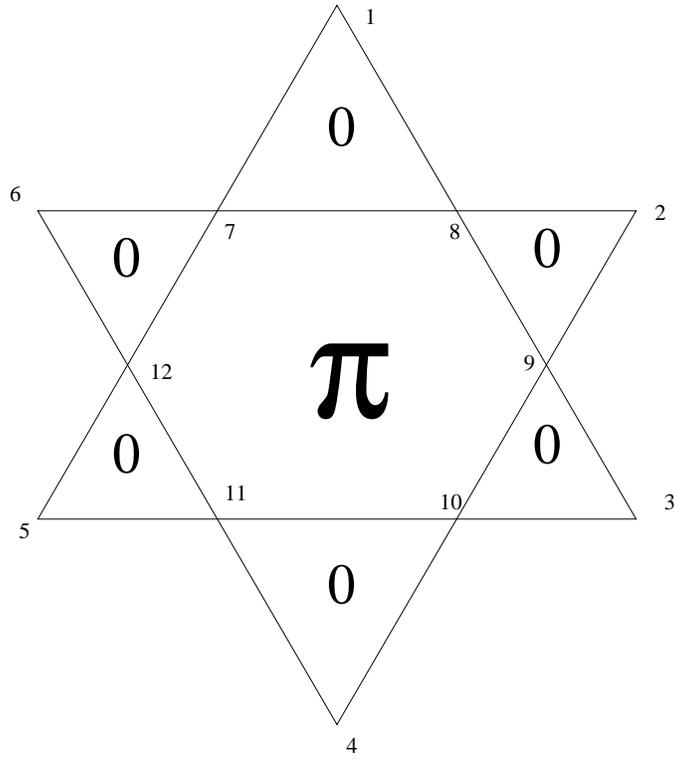,height=10cm,angle=0}
	\end{center}
	\caption{12-site unit cell, with fluxes indicated.  Small
	numbers are used to label points for reference.}
	\label{fig1}
	\end{figure}
	\begin{figure}[!t]
	\begin{center}
	\leavevmode
	\epsfig{figure=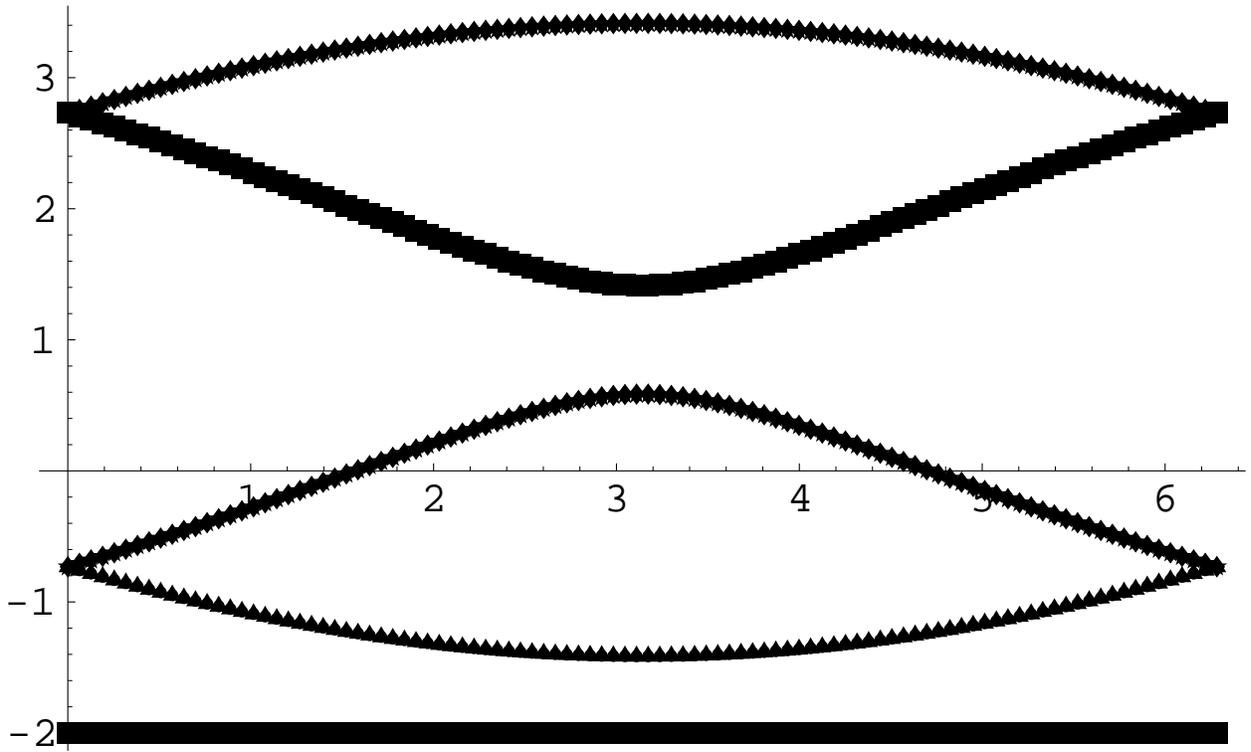,height=10cm,angle=0}
	\end{center}
	\caption{Band structure for the parent state.  We scan
along varying momenta in the $\hat x$ direction, at vanishing momentum
in the $\frac{\hat x}{2}+\frac{\sqrt{3} \hat y}{2}$ direction.}
	\label{fig2}
	\end{figure}
	\begin{figure}[!t]
	\begin{center}
	\leavevmode
	\epsfig{figure=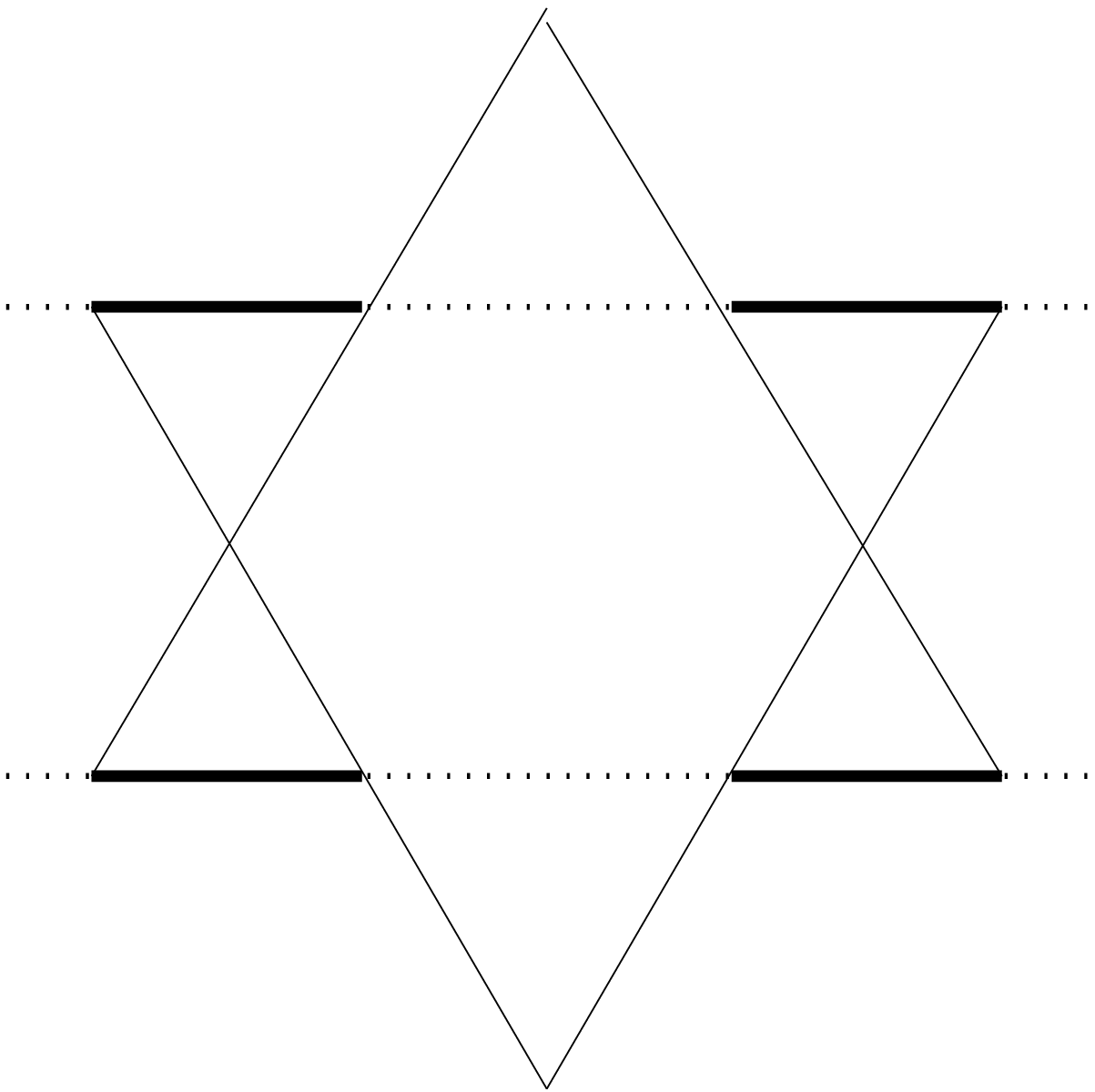,height=10cm,angle=0}
	\end{center}
	\caption{Mass perturbation to produce $M_1$.  Bold lines are
	increased in strength, dotted lines are reduced in strength.}
	\label{fig3}
	\end{figure}
	\begin{figure}[!t]
	\begin{center}
	\leavevmode
	\epsfig{figure=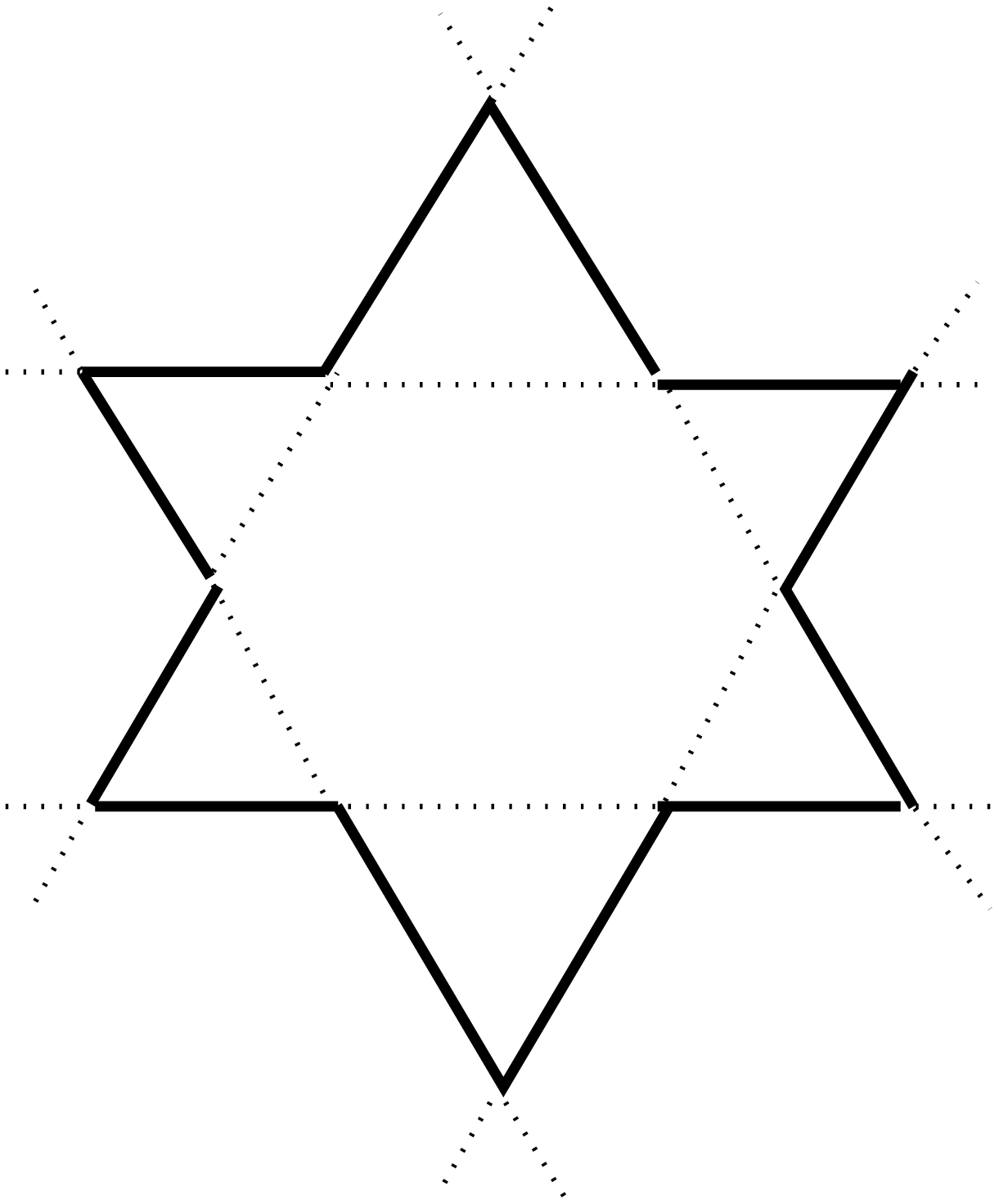,height=10cm,angle=0}
	\end{center}
	\caption{Mass perturbation to produce $M_{12}$.  Bold lines are
	increased in strength, dotted lines are reduced in strength.}
	\label{fig4}
	\end{figure}
	\begin{figure}[!t]
	\begin{center}
	\leavevmode
	\epsfig{figure=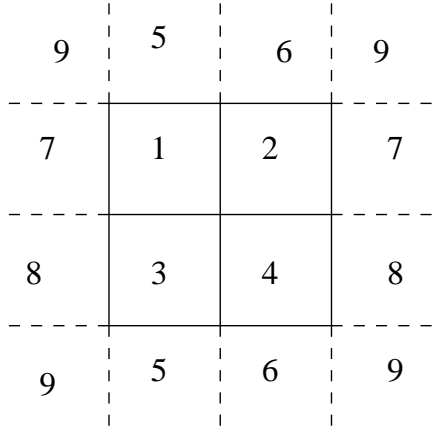,height=8cm,angle=0,scale=.75}
	\end{center}
	\caption{$\pi$-flux phase on square lattice with odd number of
sites.  Numbers label different squares, each containing
$\pi\pm\frac{\pi}{9}$ flux.}
	\label{fig5}
	\end{figure}
	\begin{figure}[!t]
	\begin{center}
	\leavevmode
	\epsfig{figure=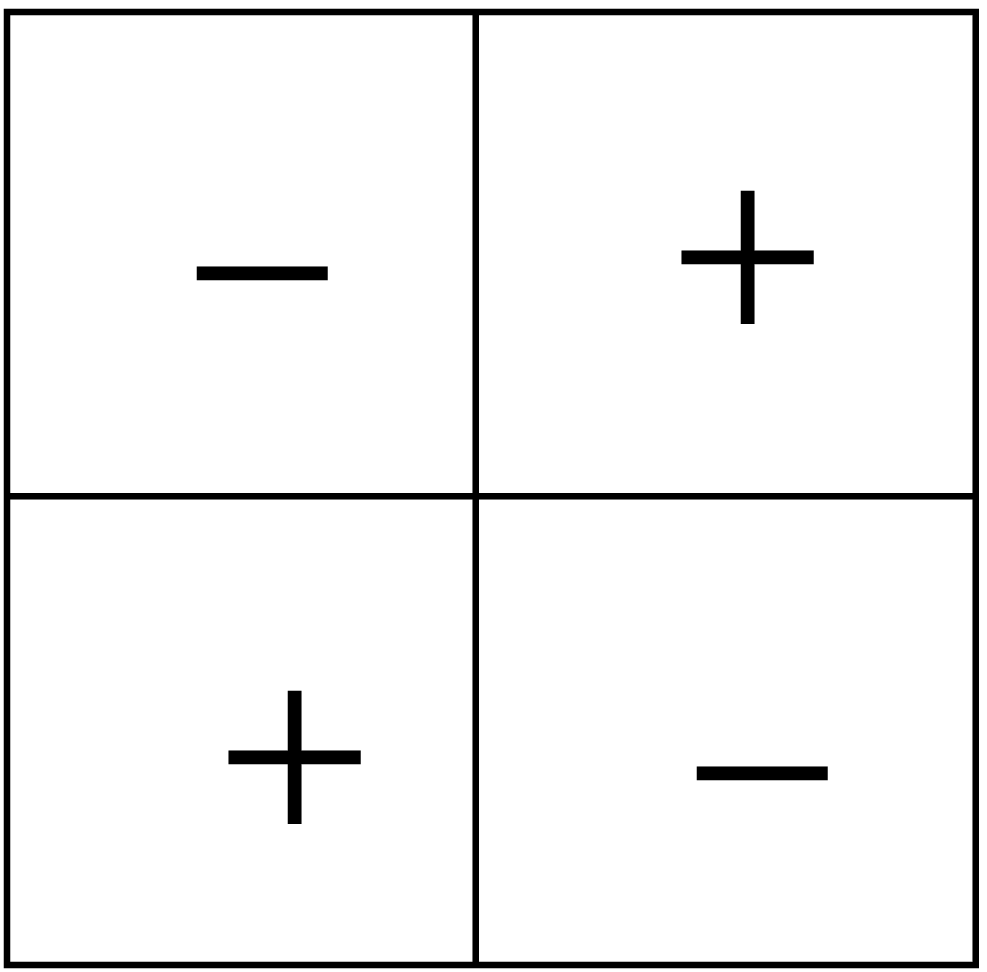,height=8cm,angle=0,scale=.75}
	\end{center}
	\caption{Enlarged torus to compute wavefunctions.}
	\label{fig6}
	\end{figure}

\begin{thebibliography}{99}
\bibitem{zengel} C. Zeng and V. Elser, Phys. Rev. B {\bf 42}, 8436 (1990).

\bibitem{square} E. Manousakis. Rev. Mod. Phys. {\bf 63}, 1 (1991) and 
references therein.

\bibitem{triangle} D. A. Huse and V. Elser, Phys. Rev. Lett {\bf 60}, 2531
(1988).

\bibitem{triangle2} T. Jolicoeur and J. C. Le Guillou, Phys. Rev. B 
{\bf 40}, 2727 (1989).

\bibitem{triangle3} L. Capriotti et. al., Phys. Rev. Lett. {\bf 82}, 3899
(1999).

\bibitem{numerics2} P. W. Leung and V. Elser, Phys. Rev. B {\bf 47},
5459 (1993).

\bibitem{numerics4} P. Lecheminant et. al., Phys. Rev. B {\bf 56}, 2521 (1997).

\bibitem{iron} M. G. Townsend, G. Longworth, E. Roudaut,
Phys. Rev. B {\bf 33} 4919 (1986).

\bibitem{iron2} S. H. Lee et. al., cond-mat/9705014.

\bibitem{jaros} A. S. Wills, A. Harrison, S. A. M. Mentink, T.
E. Mason, and Z. Tun, cond-mat/9607106.

\bibitem{scgo} X. Obradors et. al., Sol. St. Comm. {\bf 65}, 189 (1988).

\bibitem{ramirez} A. P. Ramirez, ``Geometrical Frustration", to appear
in Handbook of Magnetism.

\bibitem{mz} J. B. Marston and C. Zeng, J. Appl. Phys. {\bf 69}, 5962 (1991).

\bibitem{subir} S. Sachdev, Phys. Rev. B {\bf 45}, 12377 (1992).

\bibitem{hsus} T. C. Hsu and A. J. Schofield, J. Phys.-Cond. Matt.
{\bf 3}:(41), 8067 (1991).

\bibitem{srrvb} F. Mambrini and F. Mila, preprint cond-mat/0003080.

\bibitem{srrvb2} V. Elser, Phys. Rev. Lett. {\bf 62}, 2405 (1989).

\bibitem{srrvb3} F. Mila, Phys. Rev. Lett. {\bf 81}, 2356 (1998).

\bibitem{rvb} G. Baskaran, Z. Zou, and P. W. Anderson, Solid State
Commun. {\bf 63}, 973 (1987). 

\bibitem{htc} P. W. Anderson, Science {\bf 235}, 1196 (1987).

\bibitem{dimerize} D. S. Rokshar, Phys. Rev. B {\bf 42}, 2526 (1990).

\bibitem{ky} K. Yang, L. K. Warman, and S. M. Girvin, Phys. Rev.
Lett. {\bf 70}, 2641 (1993).

\bibitem{kl} V. Kalmeyer and R. B. Laughlin, Phys. Rev. Lett. {\bf 59},
2095 (1987).

\bibitem{numerics1}  Ch. Waldtmann, et. al.,
Eur. Phys. J. B {\bf 2}, 501 (1998).

\bibitem{mudry1} C. Mudry and E. Fradkin, Phys. Rev. B {\bf 49}, 5200
(1994).

\bibitem{sqd} S. Sachdev and M. Vojta, cond-mat/9910231.

\bibitem{hsu} T. C. Hsu, Phys. Rev. B {\bf 41}, 11379 (1990).

\bibitem{muthu} V. N. Muthukumar, private communication.

\bibitem{rules} D. S. Rokshar, Phys. Rev. Lett. {\bf 65}, 1506 (1990).

\bibitem{sun} J. B. Marston and I. Affleck, Phys. Rev. B {\bf 39} 11538
(1989).

\bibitem{polyakov} A. M. Polyakov, Gauge Fields and Strings
(Harwood Academic Publishers, New York, 1987).

\bibitem{duncan} F. D. M. Haldane and D. P. Arovas, Phys. Rev. B {\bf 52},
4223(1995).

\bibitem{cav} On the square lattice the system does not need to form the
$\pi$-flux states.  It can also form a d-wave superconducting state,
which is not equivalent to the $\pi$-flux for odd size lattices.  The
d-wave state will not give rise to a Chern number.  However, the
$\pi$-flux state on the square lattice is still a nice example to consider.

\bibitem{duncan2} F. D. M. Haldane, Phys. Rev. Lett. {\bf 61}, 2015 (1985).

\bibitem{tower} P. W. Anderson, Phys. Rev.  {\bf 86}, 694 (1952).

\bibitem{numerics3} B. Bernu, P. Lecheminant, C. Lhuillier, and
L. Pierre, Phys. Rev. B {\bf 50}, 10048 (1994).

\bibitem{path} D. P. Arovas and A. Auerbach, Phys. Rev. B {\bf 38}, 316
(1988).

\bibitem{mudry2} C. Mudry and E. Fradkin, Phys. Rev. B {\bf 50}, 11409
(1994).

\bibitem{patrick} D. H. Kim and P. A. Lee, Annals. Phys. {\bf 272}, 130
(1999).
\end{thebibliography}
\end{document}